\documentstyle[11pt,IAUS215,twoside,epsf]{article}

\def\edcomment#1{\iffalse\marginpar{\raggedright\sl#1\/}\else\relax\fi}
\reversemarginpar

\begin{document}
\vspace*{1cm}
\title{The Enigma of AA\,Dor\index{AA Dor}}
 \author{Thomas Rauch}
\affil{Dr.-Remeis-Sternwarte, D-96049 Bamberg, Germany\\
       Institut f\"ur Astronomie und Astrophysik, D-72076 T\"ubingen, Germany}
\author{Klaus Werner}
\affil{Institut f\"ur Astronomie und Astrophysik, D-72076 T\"ubingen, Germany}

\begin{abstract}
AA\,Dor\index{AA Dor} (LB\,3459\index{LB 3459}) is an eclipsing, close binary ($P = 0.26\,\mathrm{d}$)
consisting of a sdOB primary star and an unseen secondary with an extraordinary small mass.
The secondary is possibly a former planet which may have survived a common-envelope phase
and has even gained mass.

In order to investigate on an existing discrepancy between the components' mass derived
from NLTE spectral analysis and subsequent comparison to evolutionary tracks and 
masses derived from radial-velocity and the eclipse curves, we performed
phase-resolved high-resolution and high-SN spectroscopy with the UVES attached to
the ESO VLT. From the obtained spectra, we have determined AA\,Dor\index{AA Dor}'s orbital parameters 
($P = 22\,600.702 \pm 0.005 \mathrm{sec}$, $A = 39.19 \pm 0.05\mathrm{km/sec}$, $T_0 = 2451917.152690$) and the
rotational velocity ($v_{\mathrm rot} = 47 \pm 5\mathrm{km/sec}$) of the primary. 
\end{abstract}

Hilditch et al\@. (1996) have analyzed that the primary component of AA\,Dor\index{AA Dor} has a mass of
$M_1 = 0.5 M_\odot$ and the cool secondary has a very small mass of $M_2 = 0.086 M_\odot$.
The latter is in excellent agreement with lowest mass ZAMS models of Dorman et al\@. (1989).
In a recent spectral analysis (Rauch 2000) of the primary, based on high-resolution
CASPEC and IUE spectra, $M_1 = 0.330 M_\odot$ has been derived from
comparison of $T_{\mathrm{eff}}$ and $\log g$ to evolutionary tracks of post-RGB stars (Driebe et al\@. 1998) and
$M_2 = 0.066 M_\odot$ has subsequently been calculated from the system's mass function.
Thus, the secondary is possibly a planet which may have survived a common envelope phase
and has even gained mass (``late case B mass transfer'', Iben \& Livio 1993).
However, a reason for the discrepancy between these masses and those   
derived from the radial-velocity and the eclipse curves (Kilkenny et al\@. 1979) is not known.
Two possible reasons are the inaccuracy of the theoretical models by Driebe et al\@. (1998) for
AA\,Dor\index{AA Dor} and the error range for the photospheric parameter determination in Rauch (2000).
This analysis was hampered by the long exposure
times (some hours) and hence, a large orbital velocity coverage.

In order to make progress and to minimize the effects of orbital motion,
105 UVES (attached to the ESO VLT) spectra had been taken on Jan 8, 2001 with an exposure time of 180\,sec each. We achieved
a resolving power of $48\,000$ at an average $S/N\approx 20$. The spectra cover a complete orbital period of AA\,Dor\index{AA Dor}.

The phase-dependent radial velocity is determined by fitting Lorentzians
to the sharpest line in the optical spectrum of AA\,Dor\index{AA Dor}, He\,{\sc ii} $\lambda$\,4686\,\AA\ (Rauch 2000). 
For this procedure the smearing due to orbital motion during the exposure was neglected because the exposure
time is very short compared to the complete period. Then, the derived velocity curve
was fitted by a sine curve (Fig\@. 1).
We adopt $P = 22\,600.702 \mathrm{sec}$ and $A = 39.19 \mathrm{km/sec}$.
The period is by 0.02\% longer than that given by Kilkenny et al\@. 1991 ($22\,597.0 \mathrm{sec}$ from photometric data of
27 eclipses). A higher radial-velocity amplitude of $A = 40.8 \mathrm{km/sec}$ has been measured by Hilditch et al\@. (1996),
also from observations of  He\,{\sc ii} $\lambda$\,4686\,\AA.

We calculated a grid of H+He composed NLTE model atmospheres 
and performed a $\chi^2$ test (wavelength, flux level, He/H, and $v_{\mathrm{rot}}$)
in order to determine the rotational velocity of the primary by comparison to
the co-added UVES spectrum. We keep $T_{\mathrm eff} = 42 \mathrm{kK}$ and $\log g = 5.2 \,\mathrm{(cgs)}$
fixed (Rauch 2000) and achieve the best fit at $v_{\mathrm rot} = 47 \mathrm{km/sec}$ and $\mathrm{He/H} = 0.0008$ (by number).

The determined rotational velocity of the primary of $v_{\mathrm rot} = 47 \mathrm{km/sec}$
suggests, that AA\,Dor\index{AA Dor} performs a bound rotation ($45.7 \mathrm{km/sec}$). The components' mass 
determined by Rauch (2000) are verified.
The cool component has almost the same radius like Jupiter but its mass is about 1/15 M$_\odot$ which is 70 times
higher than Jupiter's mass. Thus, it lies formally within the brown-dwarf mass range ($0.013 - 0.08 M_\odot$).
However, the idea of Rauch (2000), that it may have been a planet which has survived the common envelope
phase and even has gained mass, needs further numerical simulations of the common envelope phase
to be verified.
\newline
This research was supported by the DLR (grants 50\,OR\,9705\,5 and 50\,OR\,0201).

\begin{figure}
\plotone{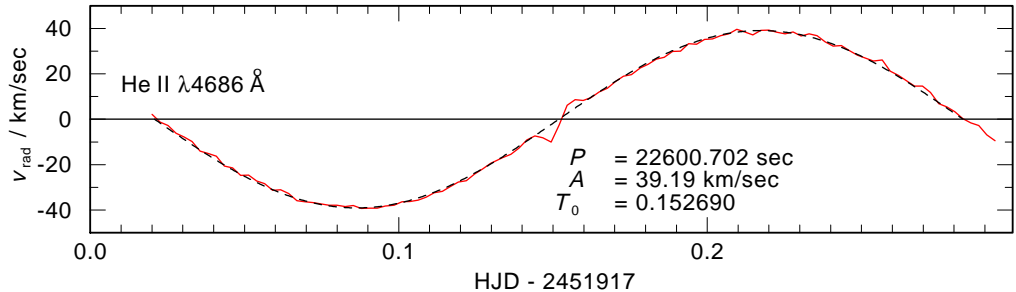}
\caption{Radial-velocity curve of AA\,Dor\index{AA Dor} measured from He\,{\sc ii} $\lambda$4686\AA\
         compared to a sine curve (dashed).
         Note the velocity jumps close to $T_0$ which are the result of the transit
         of the cool companion (Rossiter Effect, Rossiter 1924).
}
\end{figure}

\end{document}